\documentclass{iau} 
\usepackage{graphicx}

\title[Galactic globular clusters: a new catalog] 
{Galactic globular clusters: a new catalog of masses, structural parameters, velocity 
dispersion profiles, proper motions and space orbits}

\author[Hilker, Baumgardt, Sollima \& Bellini]   
{Michael Hilker$^1$
 \and Holger Baumgardt$^2$
 \and Antonio Sollima$^3$
 \and Andrea Bellini$^4$}

\affiliation{$^1$European Southern Observatory \\ Karl-Schwarzschild-Str. 2, D-85748
Garching, Germany \\ email: {\tt mhilker@eso.org} \\[\affilskip]
$^2$School of Mathematics and Physics, The University of Queensland \\ St. Lucia, 
QLD 4072, Australia \\email: {\tt h.baumgardt@uq.edu.au} \\[\affilskip]
$^3$INAF Osservatorio Astronomico di Bologna \\ via Gobetti 93/3, I-Bologna 40129, 
Italy \\ email: {\tt antonio.sollima@inaf.it} \\[\affilskip]
$^4$4Space Telescope Science Institute \\ 3700 San Martin Drive, Baltimore, MD 21218,
USA \\ email: {\tt bellini@stsci.edu}}

\pubyear{2019}
\volume{351}  
\jname{Star Clusters: From the Milky Way to the Early Universe}
\editors{A. Bragaglia, M.B. Davies, A. Sills \& E. Vesperini, eds.}
\begin{document}

\maketitle

\begin{abstract}
We collected radial velocities of more than 50.000 individual stars in 156 Galactic
globular clusters (GGC) and matched them with HST photometry and Gaia DR2 proper
motions. This allowed us to derive the GGC's mean proper motions and space velocities.
By fitting a large set of N-body simulations to their velocity dispersion and surface density 
profiles, combined with new measurements of their internal radially dependent mass 
functions, we have determined their present-day masses and structural parameters, and
for 144 GGCs their internal kinematics. We also derive the initial cluster masses by
calculating the cluster orbits backwards in time applying suitable recipes to account for 
mass-loss and dynamical friction. The new fundamental parameters of GGCs are
publicly available via an online database, which will regularly be updated.
\end{abstract}

\begin{figure}[h!]
\vspace*{-0.2 cm}
\begin{center}
\includegraphics[height=4.2cm]{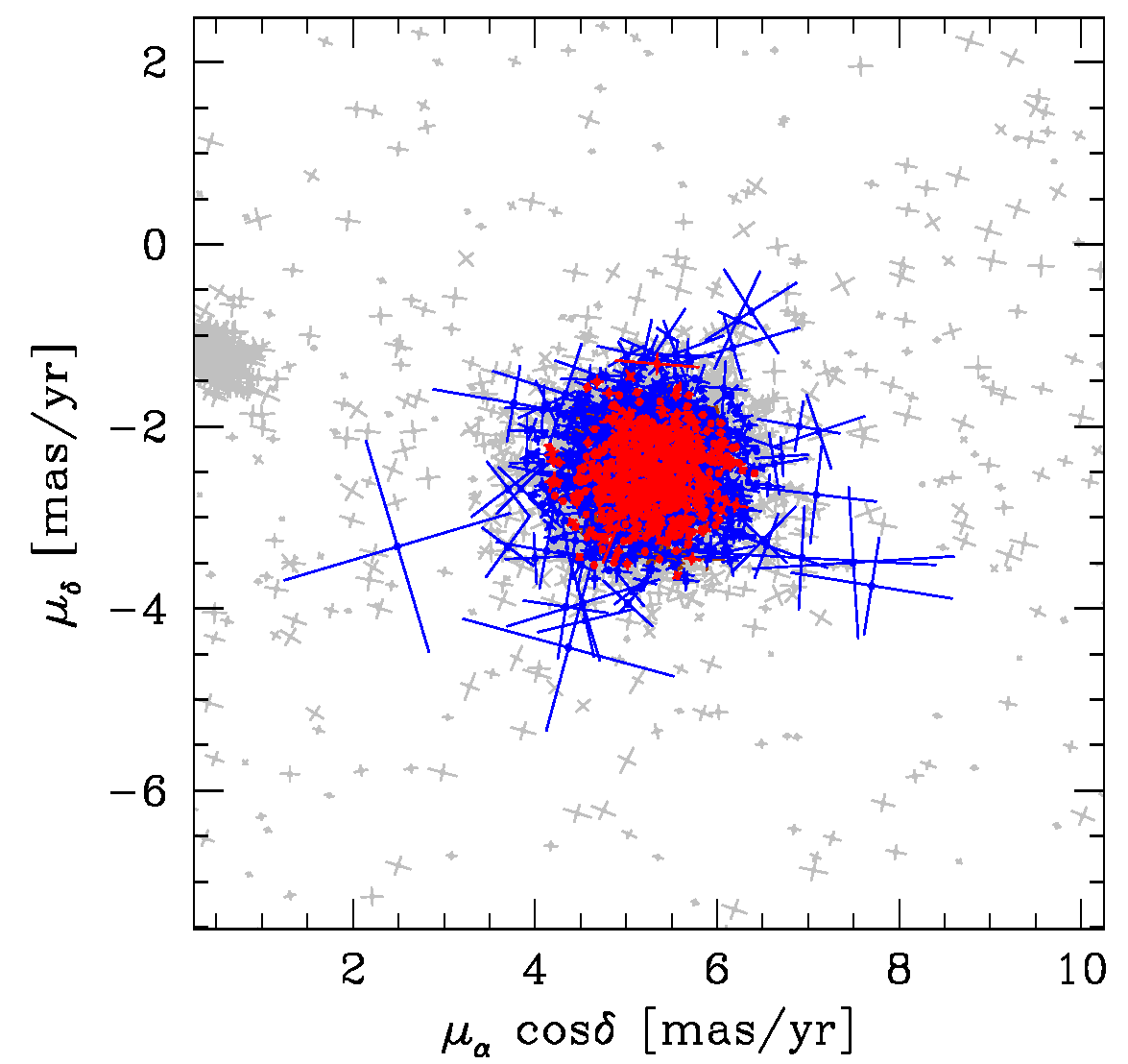}
\includegraphics[height=4.4cm]{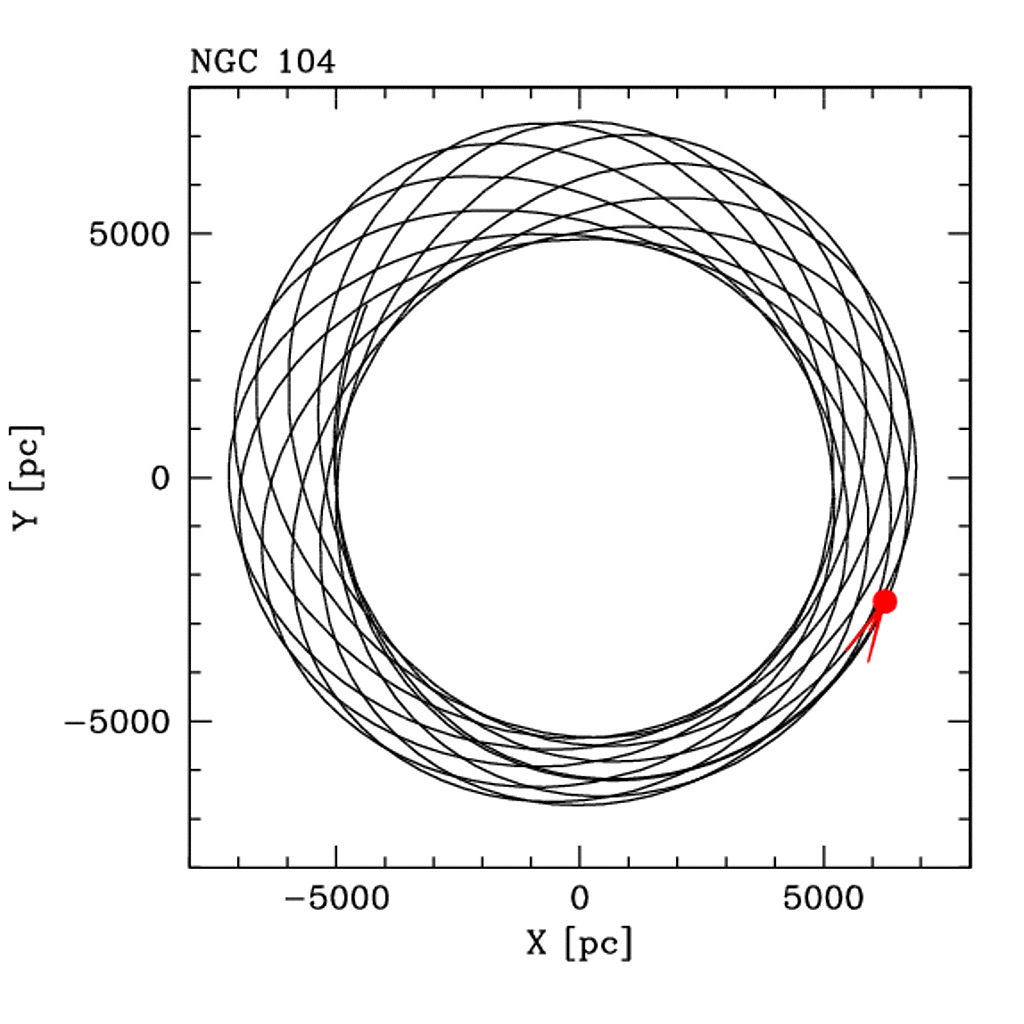} 
\includegraphics[height=4.4cm]{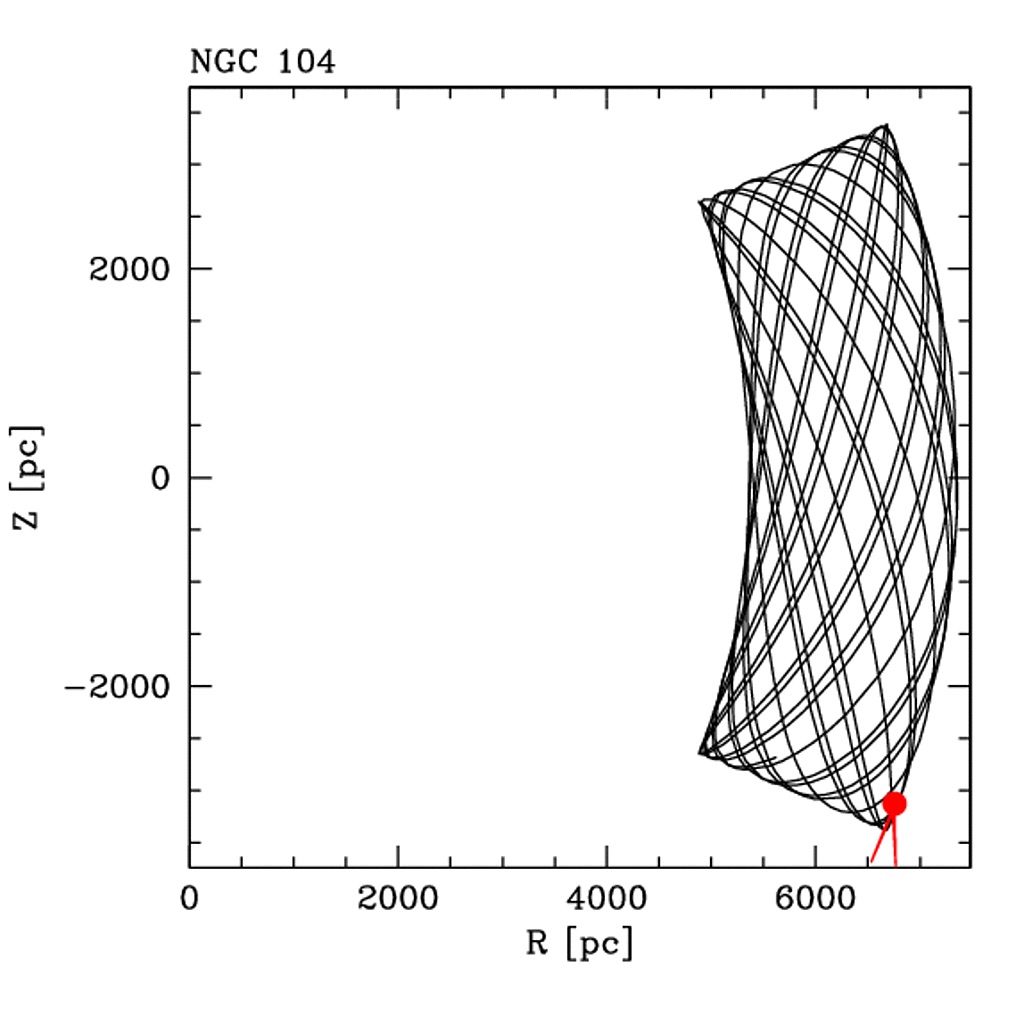}
\includegraphics[height=4.4cm]{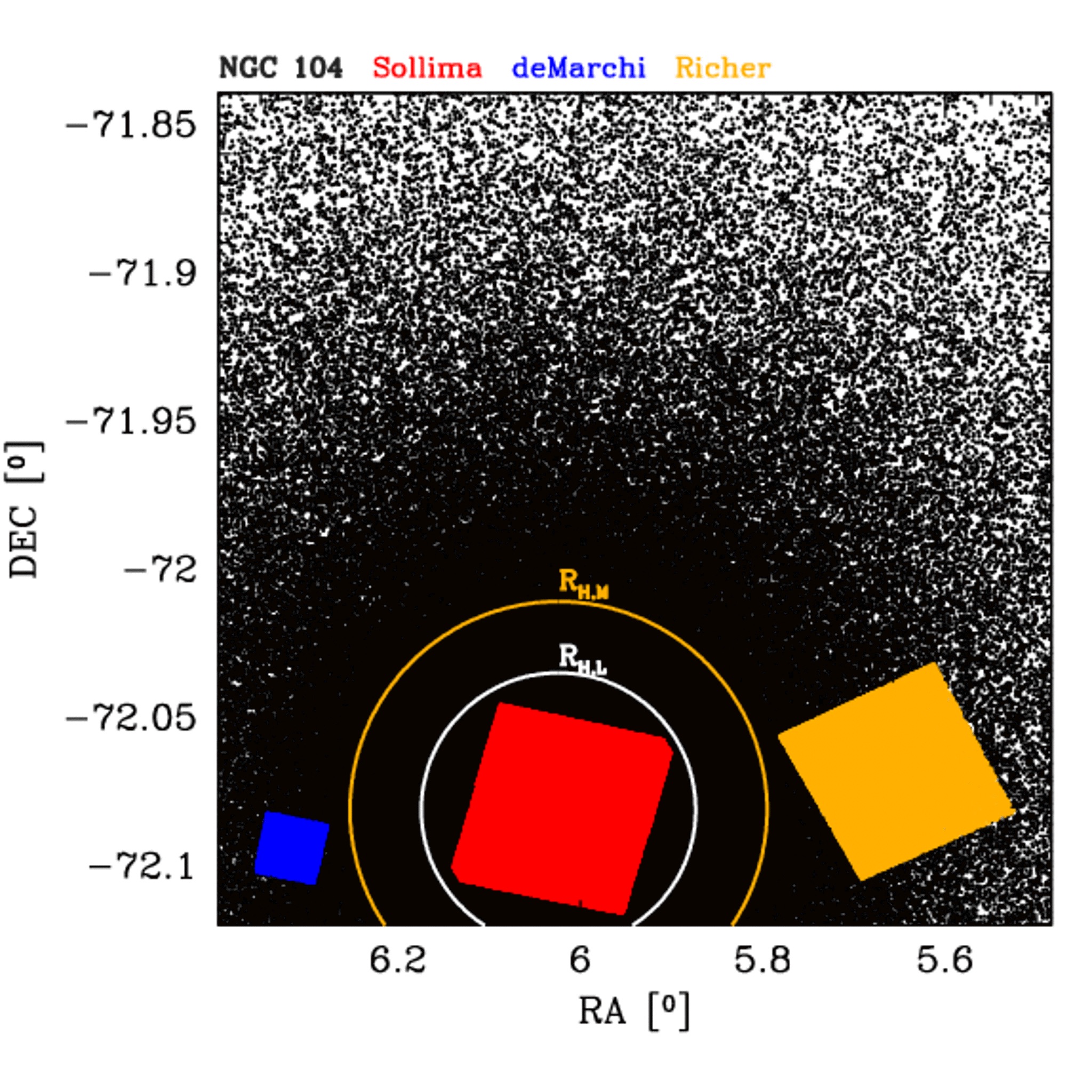} 
\includegraphics[height=4.4cm]{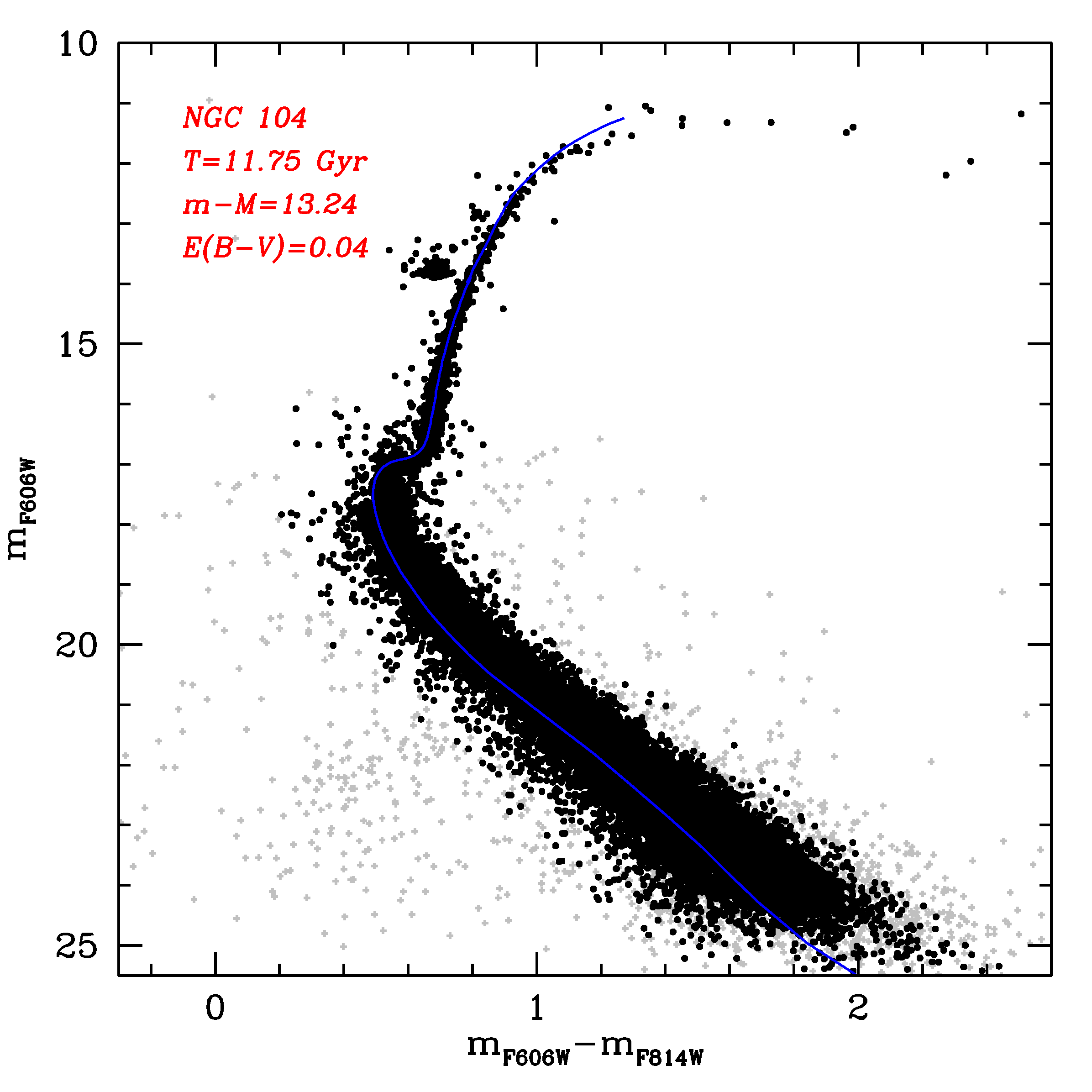} 
\includegraphics[height=4.3cm]{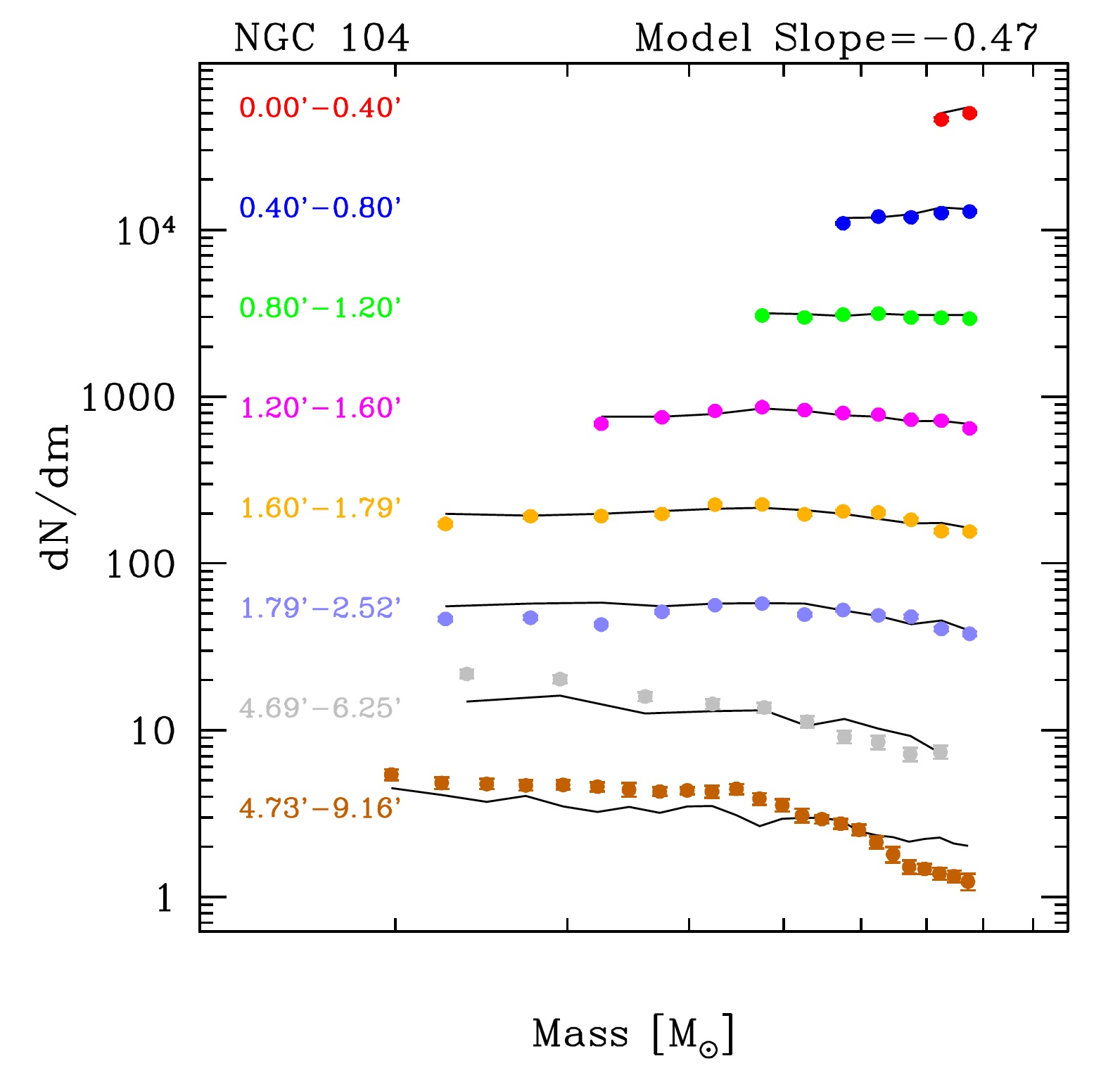} 
\includegraphics[width=13.4cm]{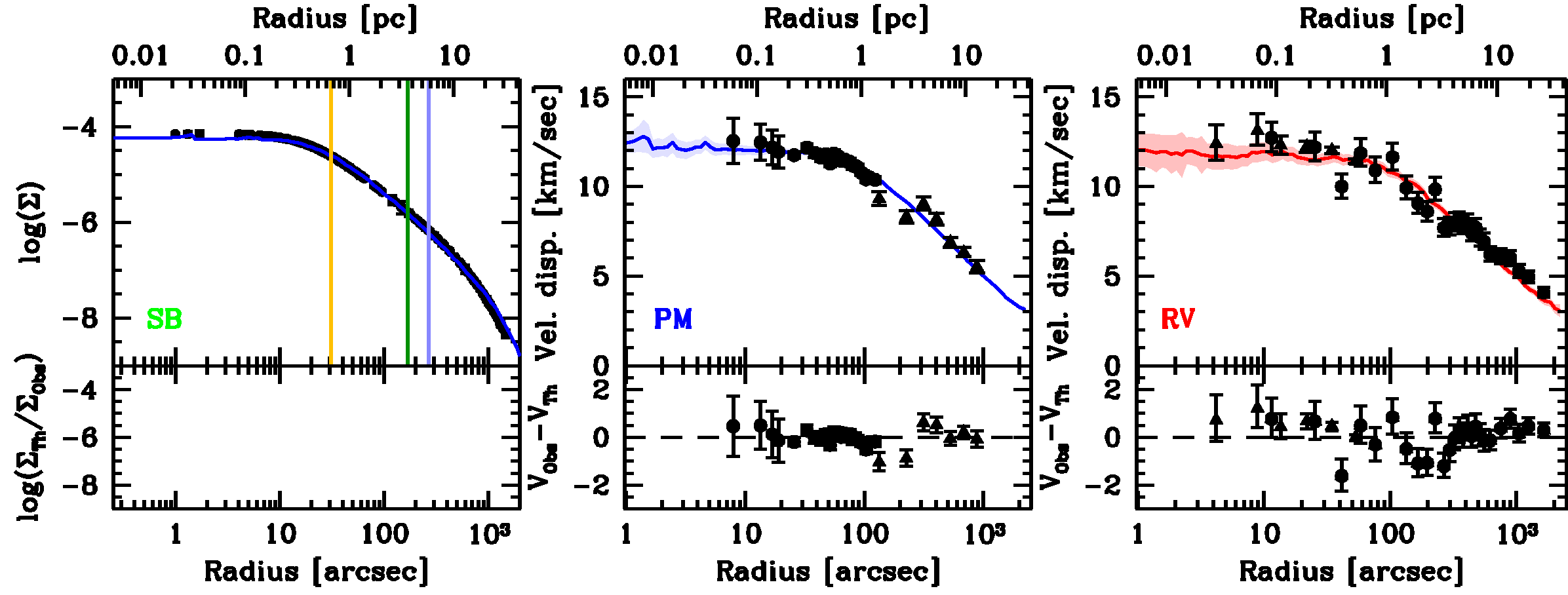} 
\caption{Example set of graphs for NGC\,104 (47 Tuc) from our catalog website. From 
top to bottom, left to right, are shown: the GC member selection from Gaia DR2
proper motions and radial velocities (red); the orbits over the last 2 Gyr in XY and
RZ Galactic coordinates; the HST observations used for the mass function analysis; the 
HST corresponding color-magnitude diagram;  mass function at various radii compared
to model fits; N-body model fits to the surface density, radial and proper motion velocity
dispersion profiles.}
\label{fig1}
\end{center}
\end{figure}

\section{The new catalog for Galactic globular clusters}

In our poster contribution we presented a new catalog of fundamental properties of 
Galactic globular clusters based on results from a series of papers: 

\cite{Baumgardt17} first compared a large grid of 900 N-body models to the velocity
dispersion and surface brightness profiles of 50 GGCs in order to determine their
masses and mass-to-light ratios. 
Additionally, \cite{Sollima17} presented the global mass functions of 35 GGCs based on 
deep HST photometry in combination with multimass dynamical models.
One year later, \cite{Baumgardt18} determined masses, stellar mass functions, and 
structural parameters of 112 GGCs by fitting a large set of N-body simulations to their 
velocity  dispersion and surface density profiles. The velocity dispersion profiles were 
calculated based on a combination of more than 15.000 high-precision radial velocities, 
which were derived from archival ESO/VLT and Keck spectra, together with
$\sim$20.000 published  radial velocities from the literature. 
When Gaia DR2 became public, \cite{Baumgardt19} presented mean proper motions
and space velocities of 154 GGCs and the velocity  dispersion profiles of 141 globular 
clusters based on a combination of Gaia DR2 proper motions with ground-based
line-of-sight velocities. The combination of these velocity dispersion profiles with new 
measurements of the internal mass functions allowed to model the internal kinematics
of 144 GGCs, more than 90 per cent of the currently known Milky Way globular cluster 
population. 
Finally, \cite{Sollima19} analysed the internal kinematics of 62 GGCs, finding significant
rotation in 15 of them.

All those results are combined in our new GGC catalog. The online version of this
catalog, which is and always will be the reference for the most up-to-date results of our
work, can be accessed under the following link: 

https://people.smp.uq.edu.au/HolgerBaumgardt/globular/

\noindent
There we provide several tables:
\begin{itemize}
\item A table containing the mean radial velocities, proper motions and orbital parameters     
of 156 GGCs, derived from the GAIA proper motions and our new radial velocities (\cite[Baumgardt \& Hilker 2018, Baumgardt et al. 2019]{Baumgardt18, Baumgardt19}). The orbital integrations were done in the \cite{Irrgang13} Milky Way mass model.
\item A table containing the masses and structural parameters of 156 GGCs, based on our N-body fits (\cite[Baumgardt \& Hilker 2018]{Baumgardt18}).
\item Velocity dispersion profiles of 139 GGCs. They are based on the following works: \cite{Watkins15}, \cite{Baumgardt17}, \cite{Kamann18}, \cite{Baumgardt18} and \cite{Baumgardt19}.
\item Individual radial velocities for more than 50.000 stars in 122 GGCs, derived from ESO proposals prior to 2014. The data files also contain the Gaia DR2, APOGEE DR14, Keck/DEIMOS, Keck/HIRES and Keck/NIRSPEC radial velocities which are not included in the MNRAS paper.
\item For 156 GGCs: figures on the Gaia selection, HST photometry, orbit over the last 2 Gyr, color magnitude diagram, mass function at different radii, N-body fits to the surface density and velocity dispersion profiles, see an example in Fig.\,\ref{fig1}.
\end{itemize}
Regular updates of those tables including new radial velocity and photometry data as well as model fits are planned.
 
\section{Some key results}

The compilation and calculation of newly derived structural and dynamical parameters
of almost all Milky Way GCs allowed us to investigate several global properties of the 
Galactic globular cluster system and to derive fundamental correlations. Our most 
prominent findings are listed in the following.

{\underline{\it Present-day mass function slope}}. There exists a strong correlation 
between the stellar mass function (MF) of a GC and the amount of mass lost from the
cluster as well as their relaxation time \cite[(Sollima \& Baumgardt 2017, Baumgardt \&
Hilker 2018)]{Sollima17, Baumgardt18}. Both, metal-rich and metal-poor GGCs, follow
the same correlation. Dynamical evolution is the main mechanism shaping the MF of
stars in clusters. There is a strong hint that GGCs are formed with a bottom-light IMF.
The present-day MF slope anti-correlates with the half-mass relaxation time, while the
fraction of rotational kinetic energy correlates with it \cite[(Baumgardt et al. 2019)]{Baumgardt19}.

{\underline{\it Initial GC masses}}. The surviving Galactic GC population has lost about 
80\% of its initial mass: at formation, GCs were about 5x more massive. If GCs started 
from a log-normal mass function, the MW contained $\sim$500 GCs initially, with a 
combined mass of about $2.5\times10^8 M_{\odot}$. For a power-law initial mass 
function, the initial mass in GCs could have been about 3$\times$ higher
\cite[(Baumgardt et al. 2019)]{Baumgardt19}.

{\underline{\it Fraction of first population stars}}. The fraction of first-population stars in
a GGC correlates with its central escape velocity as well as with its total present-day 
mass, but not with its global MF slope \cite[(Baumgardt \& Hilker 2018)]{Baumgardt18}, 
see Fig.\,\ref{fig3}. 
This could indicate that the ability of a globular cluster to keep the wind ejecta from the
polluting star(s) is the crucial parameter determining the presence and fraction of
second-population stars and not its later dynamical mass loss.

\begin{figure}[h!]
\begin{center}
\includegraphics[height=6.4cm]{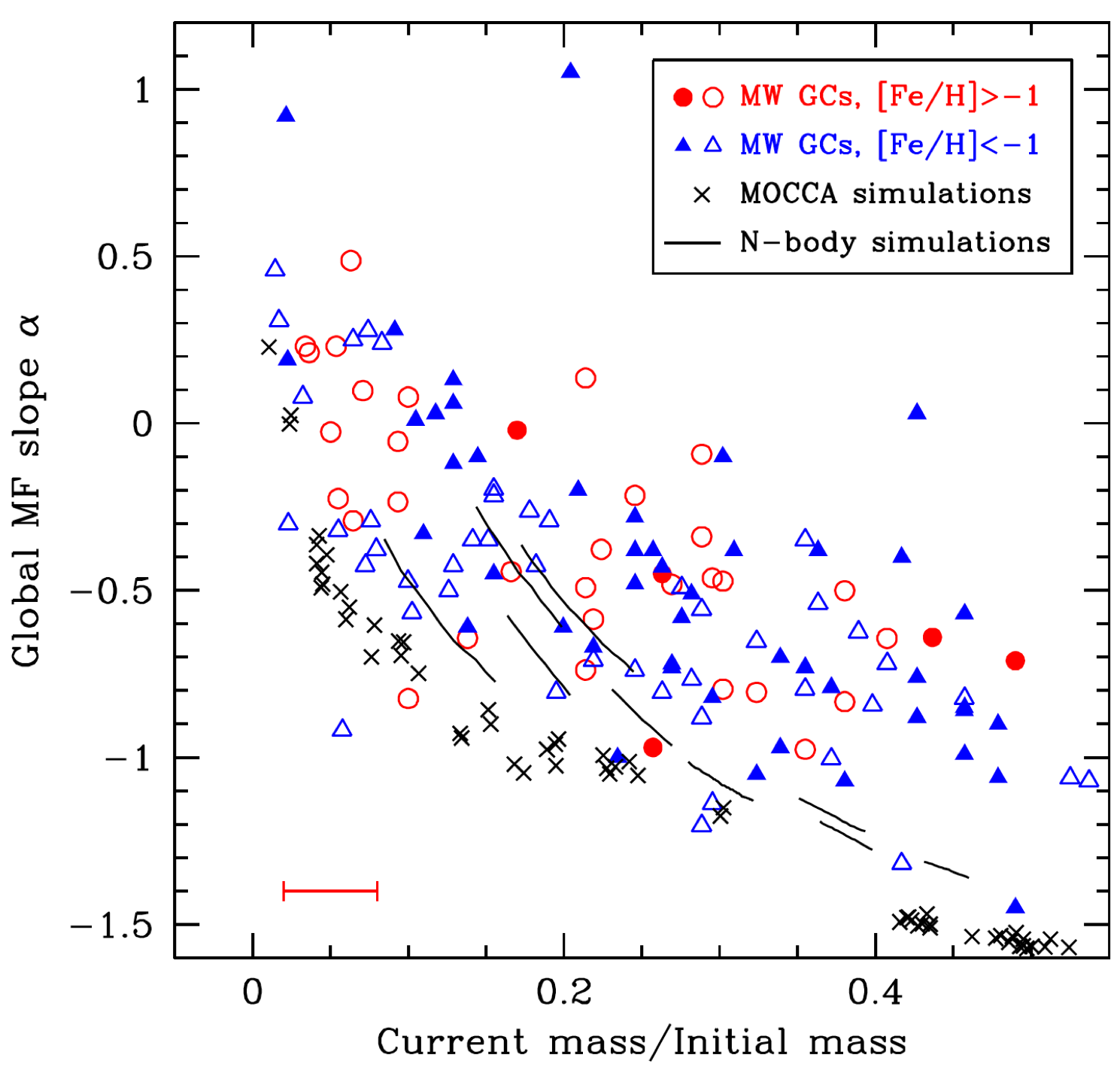} 
\includegraphics[height=6.4cm]{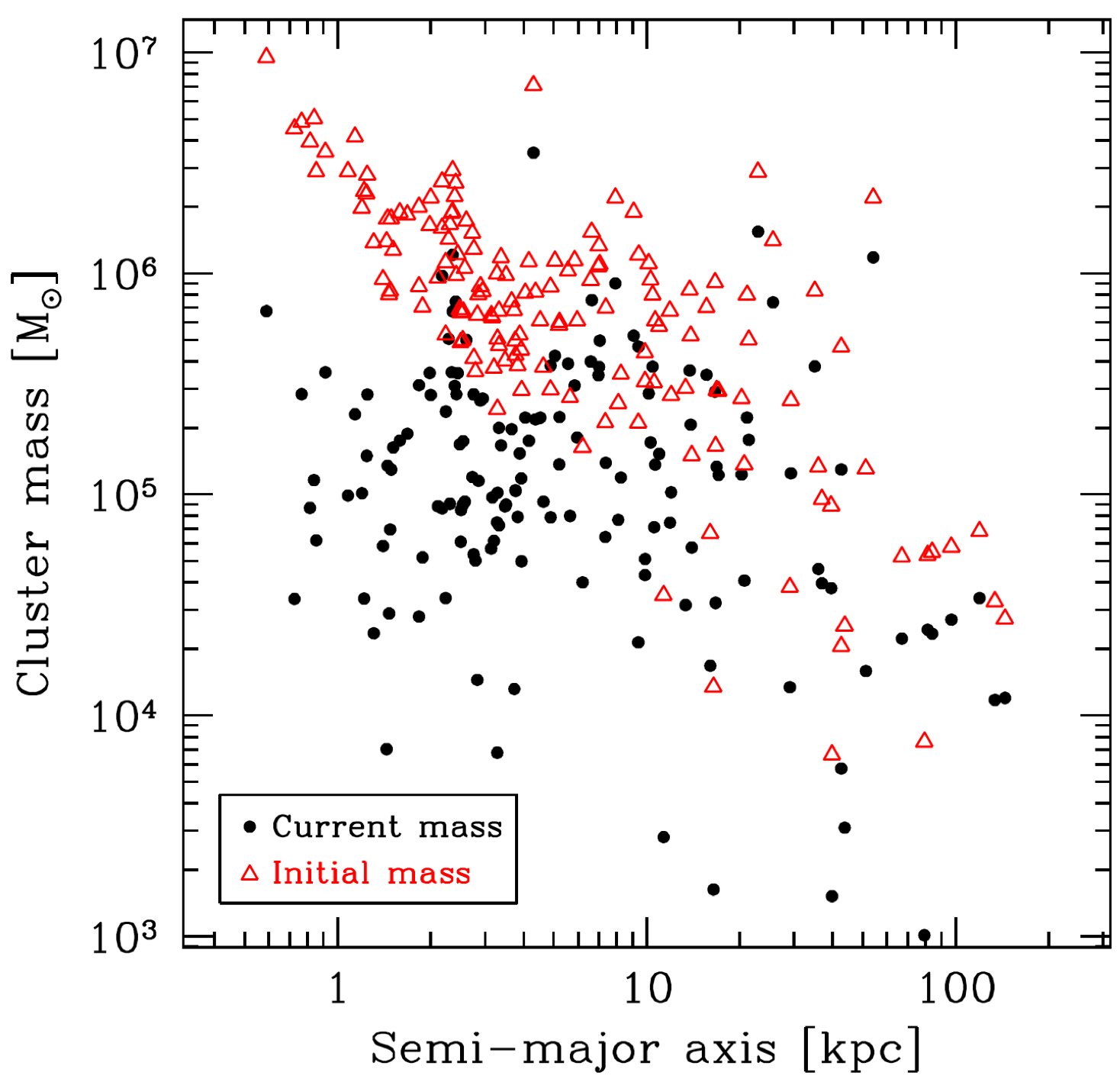}
\caption{Left: Mass fraction $M(t)/M_{\rm ini}$ remaining in a GC versus the slope of the 
best-fitting power-law to the mass function of main-sequence stars between 0.2 and 0.8
$M_{\odot}$. Right: Current and initial cluster masses as a function of the semi major
axis of a cluster?s orbit in the Milky Way. For further details see \cite{Baumgardt19}.}
\label{fig2}
\end{center}
\end{figure}

\begin{figure}[h!]
\begin{center}
\includegraphics[width=13.4cm]{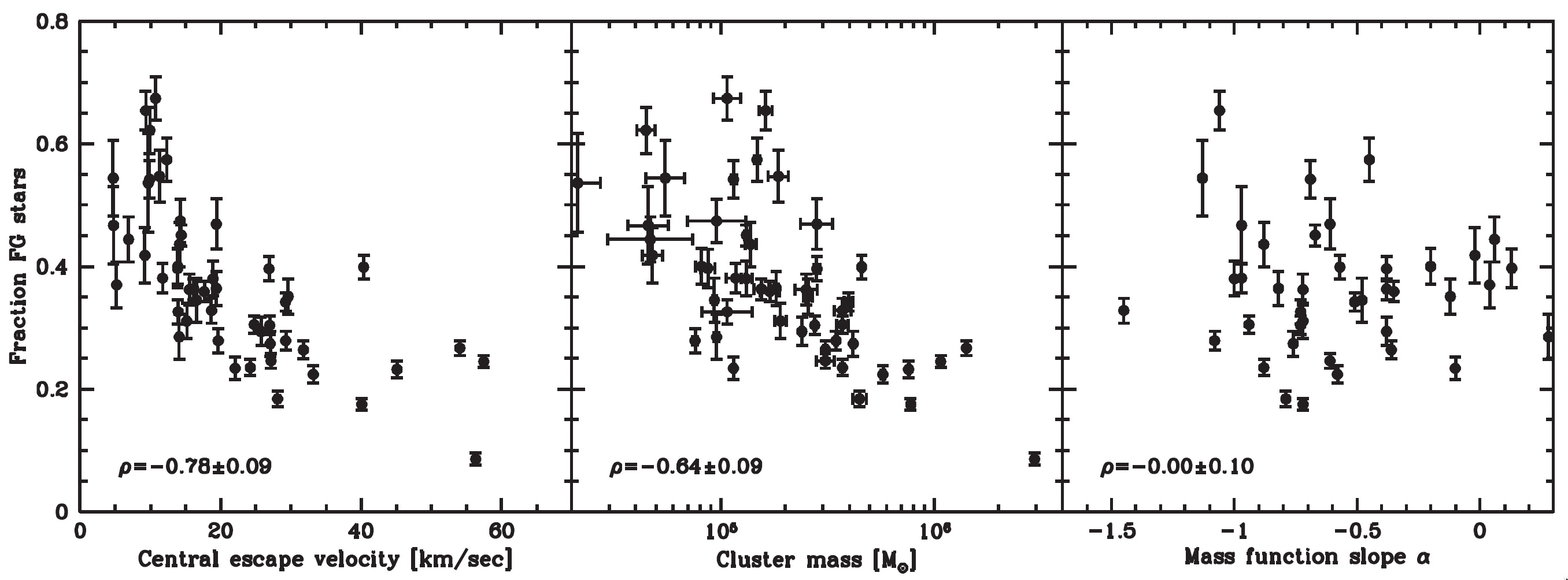} 
\caption{Correlation of the fraction of FG stars as determined by \cite{Milone17} with
the escape velocity (left), the cluster mass (middle), and the mass function slope (right),
taken from \cite{Baumgardt18}.}
\label{fig3}
\end{center}
\end{figure}

\end{document}